\documentclass[preprintnumbers,amssymb]{revtex4}
\usepackage{graphicx}
\usepackage{dcolumn}
\usepackage{bm}
\usepackage{amssymb}
\usepackage{amsmath}
\begin{document}
\addtolength{\voffset}{-.75cm}
\addtolength{\textheight}{1.5cm}
\addtolength{\hoffset}{-0.25cm}
\addtolength{\textwidth}{.5cm}
\title{ Short Baseline Neutrino Oscillations and  a New Light Gauge Boson}
\author{Ann E. Nelson$^{1,2}$ and Jonathan Walsh$^1$}
\affiliation{
$^1$Department of Physics, Box 1560, University of Washington,\\
           Seattle, WA 98195-1560, USA\\
           and\\
           $^2$Instituto de F\'isica Te\'orica UAM/CSIC\\
           Facultad de Ciencias, C-XVI\\
           Universidad Aut\'onoma de Madrid\\
           Cantoblanco, Madrid 28049, SPAIN
}
\newcommand{\geff}{g_{\rm eff}}
\newcommand{\meff}{m_{\rm eff}}
\newcommand{\p}[0]{\partial}
\renewcommand{\d}[0]{\textrm{ d}}
\newcommand{\R}[0]{\mathbb{R}}
\newcommand{\C}[0]{\mathbb{C}}
\newcommand{\bra}[1]{\big<#1\big|}
\newcommand{\ket}[1]{\big|#1\big>}
\renewcommand{\matrix}[4]{\left( \begin{array}{c c} #1 & #2 \\ #3 & #4 \end{array} \right)}
\newcommand{\pf}[2]{\frac{\partial #1}{\partial #2}}
\newcommand{\df}[2]{\frac{\textrm{d}#1}{\textrm{d}#2}}
\newcommand{\be}[0]{\begin{equation*}}
\newcommand{\ee}[0]{\end{equation*}}
\newcommand{\sfrac}[2]{\textstyle{\frac{#1}{#2}}}
\newcommand{\Res}[0]{\textrm{Res}}
\newcommand{\lc}[1]{\lambda_{#1}}
\newcommand{\del}[0]{\nabla}
\newcommand{\braket}[2]{\big<#1\big|#2\big>}
\newcommand{\beq}{\begin{eqnarray}}
\newcommand{\eeq}{\end{eqnarray}}
\newcommand{\nn}{\nonumber}
\newcommand{\bml}{ $U(1)_{\textrm{\tiny{B-L}}}$}
\newcommand{\rbml}{ \rho_{\textrm{\tiny{B-L}}}} 
\def\ltap{\ \raise.3ex\hbox{$<$\kern-.75em\lower1ex\hbox{$\sim$}}\ }
\def\gtap{\ \raise.3ex\hbox{$>$\kern-.75em\lower1ex\hbox{$\sim$}}\ }
\def\CO{{\cal O}}
\def\CL{{\cal L}}
\def\CM{{\cal M}}
\def\tr{{\rm\ Tr}}
\def\CO{{\cal O}}
\def\CL{{\cal L}}
\def\CM{{\cal M}}
\def\mpl{M_{\rm Pl}}
\newcommand{\Dslash}{D\llap{/\kern+2.1pt}}
\newcommand{\bel}[1]{\be\label{#1}}
\def\al{\alpha}
\def\bt{\beta}
\def\eps{\lambda}
\def\eg{{\it e.g.}}
\def\ie{{\it i.e.}}
\def\mn{{\mu\nu}}
\newcommand{\rep}[1]{{\bf #1}}
\def\be{\begin{equation}}
\def\ee{\end{equation}}
\def\bea{\begin{eqnarray}}
\def\eea{\end{eqnarray}}
\newcommand{\eref}[1]{(\ref{#1})}
\newcommand{\Eref}[1]{Eq.~(\ref{#1})}
\newcommand{\gsim}{ \mathop{}_{\textstyle \sim}^{\textstyle >} }
\newcommand{\lsim}{ \mathop{}_{\textstyle \sim}^{\textstyle <} }
\newcommand{\vev}[1]{ \left\langle {#1} \right\rangle }
\newcommand{\ev}{{\rm eV}}
\newcommand{\kev}{{\rm keV}}
\newcommand{\Mev}{{\rm MeV}}
\newcommand{\gev}{{\rm GeV}}
\newcommand{\tev}{{\rm TeV}}
\newcommand{\mev}{{\rm MeV}}
\newcommand{\mnu}{\ensuremath{m_\nu}}
\newcommand{\mlr}{\ensuremath{m_{lr}}}
\newcommand{\acc}{\ensuremath{{\cal A}}}
\newcommand{\mav}{MaVaNs}
\newcommand{\nusm}{$\nu$SM }

\begin{abstract}
We consider a model of neutrino oscillations  with three  additional sterile neutrinos and a   gauged $B-L$  interaction. We find allowed values of the model parameters which can reconcile the results of the evidence for $\bar{\nu}_{\mu}\rightarrow \bar{\nu}_e$ conversion seen at the LSND neutrino oscillation experiment with the null results of the MiniBooNE experiment.  A portion of the low energy excess of $\nu_e$ events seen at MiniBooNE can arise naturally, and we make a quantitative prediction for the forthcoming anti-neutrino oscillation results at MiniBooNE.  
\end{abstract}
\maketitle

\section{Introduction}
The standard three neutrino oscillation model is experimentally established  by long baseline neutrino experiments \cite{Donini:1999jc,Maltoni:2002ni,review,deGouvea:2004gd,Mohapatra:2005wg,Fogli:2005cq,Strumia:2006db,Mohapatra:2006gs,GonzalezGarcia:2007ib,Conrad:2007ea}, which are sensitive to   differences in neutrino mass  squared which are  much smaller than 0.1 eV$^2$.  The neutrino flavor eigenstates $e, \mu, \tau$ are related to the mass eigenstates via a 3-by-3 unitary matrix, neglecting a CP-violating phase $\delta$:
\be\label{stdosc}
\left( \begin{array}{c}
\nu_e \\
\nu_{\mu} \\
\nu_{\tau}
\end{array} \right) = \left( \begin{array}{c c c}
1 & 0 & 0 \\
0 & c_{23} & s_{23} \\
0 & -s_{23} & c_{23}
\end{array} \right)\left( \begin{array}{c c c}
c_{13} & 0 & s_{13} \\
0 & 1 & 0 \\
-s_{13} & 0 & c_{13}
\end{array} \right)\left( \begin{array}{c c c}
c_{12} & s_{12} & 0 \\
-s_{12} & c_{12} & 0 \\
0 & 0 & 1
\end{array} \right)\left( \begin{array}{c}
\nu_1 \\
\nu_2 \\
\nu_3
\end{array} \right)\ ,
\ee
where $c_{ij} = \cos\theta_{ij}$, $s_{ij} = \sin\theta_{ij}$.  The three angles $\theta_{ij}$ have been measured to good accuracy:
\be\label{stdangles}
\begin{split}
\tan^2 \theta_{12} &= 0.45\pm0.05 \\
\sin^2 2\theta_{13} &= 0^{+0.05}_{-0} \\
\sin^2 2\theta_{23} &= 1.0^{+0}_{-0.1}
\end{split}
\ee
The neutrino mass squared differences have also been determined:
\be\label{stdmasses}
\begin{split}
\Delta m_{12}^2 &= (8.0\pm0.3)\cdot10^{-5}\textrm{ eV}^2 \\
\big|\Delta m_{23}^2\big| &= (2.5\pm0.2)\cdot10^{-3}\textrm{ eV}^2
\end{split}
\ee
While many experiments have confirmed this picture, a pair of short baseline experiments suggest new physics may be contributing to neutrino oscillations.  The LSND experiment found evidence at the 3$\sigma$ level for $\bar{\nu}_{\mu}\rightarrow\bar{\nu}_e$ oscillations \cite{Athanassopoulos:1996wc,Athanassopoulos:1997pv,Aguilar:2001ty}.  This result does not fit into the standard oscillation model, as the LSND experiment was sensitive to oscillations between neutrinos with a mass squared difference greater than 0.1 eV$^2$. 

In an attempt to confirm or rule out the LSND results, the MiniBooNE experiment searched for $\nu_{\mu}\rightarrow\nu_e$ oscillations over the same $\Delta m^2$ range as LSND, but with higher energy and longer baseline.  The initial results of the experiment are inconsistent with a standard oscillation interpretation of LSND, but saw an excess of electron candidates in the low energy spectrum of their data \cite{AguilarArevalo:2007it}.  The low energy excess  occurs at energies below 475 MeV, in a region not included in  the  oscillation analysis, and where the uncertainty in the  background   is larger than in the analysis region.  If the excess at MiniBooNE is a  physics signal rather than some nonunderstood background, then  new physics is needed, which might also be able to  explain the results of LSND.
Several attempts have been made to reconcile the LSND and MiniBooNE results with each other and with other  oscillation experiments. This appears to  require new ingredients besides simple neutrino oscillations, even if more sterile neutrinos  and CP violation are introduced \cite{Pas:2006tk,Maltoni:2007zf,Giunti:2007xv,Harvey:2007rd,Nunokawa:2007qh,Schwetz:2007cd,Conrad:2007ea}. 

To accommodate the  results of the MiniBooNE and LSND experiments, we introduce a model for neutrino oscillations that can be compatible with the long baseline results, the various null short baseline results,  and the LSND and MiniBooNE anomalies. As in the conventional seesaw model \cite{GellMann:1976pg,Minkowski:1977sc,Mohapatra:1979ia,Yanagida:1980xy,Schechter:1980gr,Lazarides:1980nt}, neutrino masses are allowed via the introduction of 3 sterile neutrinos.  
However,  the sterile neutrinos have small masses in the  eV range. (Such ``miniseesaw'' models have been considered before, for a variety of reasons \cite{Dvali:1998qy,Langacker:1998ut,Arkani-Hamed:2000bq,deGouvea:2005er,deGouvea:2006gz,Montero:2007cd}.) Vacuum mixing angles for oscillations between the active and sterile sectors are all fairly small.  A large MSW-type potential for neutrinos in matter \cite{Mikheev:1986gs,wolfenstein}, arising from a $B-L$ gauge interaction, gives strong energy dependence in the short baseline small mixing angle oscillations.    Three sterile neutrinos are an integral part of the model, as they are required for gauge anomaly cancellation and  to allow for neutrino masses.  
Long baseline data indicates that 
new interactions for the active neutrinos should either be flavor diagonal or   weak \cite{Grifols:2003gy,Barranco:2005ps,GonzalezGarcia:2006vp}.
However because the active neutrinos all have the same $B-L$ charge, in the absence of mixing with sterile neutrinos the   $B-L$ potential has no  effect on  the  oscillations among active neutrinos.  While the $B-L$ potential does give a significant matter effect for the mixing with sterile neutrinos, when the active-sterile  mixing is small,  the  effects  of the new interaction on long baseline, large mixing angle oscillations among active neutrinos can be negligible.
   
We note that in order to get a significant matter effect at short baseline, we will need the $B-L$ boson to be   light.  We  will argue that there is a phenomenologically allowed window for a weakly coupled $U(1)_{B-L}$ gauge interaction with a light (tens of keV) vector boson with a coupling constant less than $\sim 10^{-5}$.  While the effects  of such a boson are small in the precision electroweak and QED experiments,  there can be a significant effect on  neutrino oscillations.   

Several features of the LSND and MiniBooNE experimental results arise naturally in the model. Short baseline  active flavor oscillations at higher neutrino energy are suppressed by the MSW-type potential in matter, and anti-neutrinos tend to have higher oscillation probabilities than neutrinos.  The results of MiniBooNE relative to LSND can be qualitatively explained by these features -- MiniBooNE searched for oscillations at higher energy and with neutrinos, while the strongest evidence for oscillation found at the lower energy LSND experiment was with anti-neutrinos.  Further, part of the low energy electron neutrino excess at MiniBooNE can be explained by the strong energy dependence of  oscillation parameters in this model, although it is difficult to fit the entire excess.

After discussing the model and deriving the formula for oscillations, we proceed to fit the results of LSND,  MiniBooNE, and short baseline neutrino disappearance experiments numerically.   We find a substantial portion of parameter space capable of fitting all experiments simultaneously, and give a prediction for the outcome of the MiniBooNE anti-neutrino oscillation experiment.

\section{The LSND and MiniBooNE Experiments}
We give a brief summary of the two experiments.  LSND observed a 3.2 $\sigma$ excess of $\bar{\nu}_{\mu}\rightarrow\bar{\nu}_e$ oscillation events using $\mu^+$ decays at rest, which produced a muon anti-neutrino energy spectrum up to the Michel endpoint for muon decay.  A secondary analysis looked for oscillations in $\nu_{\mu}\rightarrow\nu_e$ from $\pi^+$ decays in flight, which yield neutrino energies in the range 60 to 200 MeV.  This analysis initially observed a 2.6 $\sigma$ excess, but a later analysis showed a reduced signal of 0.6 $\sigma$.  The observed anti-neutrino oscillation signal is
\begin{equation*}
P(\bar{\nu}_{\mu}\rightarrow\bar{\nu}_e) = (0.264\pm0.081)\% \\
\end{equation*}
The LSND detector was centered 30 m from the neutrino source, meaning the anti-neutrino oscillation experiment was sensitive to a neutrino mass squared difference
\be
\Delta m^2 \sim \Big(\frac{L}{E}\Big)^{-1} = 0.1 - 0.4 \textrm{ eV}^{2}\ .
\ee
MiniBooNE was designed to be sensitive to the same range in $\Delta m^2$, but is characteristically quite different from LSND.  MiniBooNE searched for $\nu_{\mu}\rightarrow\nu_e$ oscillations using muon neutrinos with energies between 200 MeV and 3 GeV, and the detector was situated 541 m from the neutrino source.

In their recently released results, MiniBooNE found no evidence for neutrino oscillations in the energy range 475 - 3000 MeV,  but an unexplained low energy excess of $\nu_e$ in the energy range 200 MeV to 475 MeV.  The fit of the excess to conventional oscillations is poor, and the excess corresponds to percent level oscillations in this energy regime.  The result of an analysis of $\bar{\nu}_{\mu}\rightarrow\bar{\nu}_e$ oscillations at MiniBooNE is expected in the coming months, and could prove to be a strong test of the anomalies seen at both LSND and MiniBooNE.

\section{A Renormalizable Model of Neutrino Masses with a Light $B-L$ Gauge Boson}
We  motivate the model by considering the phenomenology of a flavor universal MSW-type matter effect.  We   take the results of LSND and MiniBooNE at face value: they found different oscillation results, and the   key differences between the two experiments are the energy scales and the use of neutrinos at MiniBooNE and anti-neutrinos at LSND.  We thus need an energy dependent effect and/or one which distinguishes between neutrinos and anti-neutrinos. Vector interactions with matter do both. We take this as evidence for a new gauge interaction, whose potential provides a strong energy dependence in short baseline oscillations.  In this model we utilize a weakly coupled, flavor universal $U(1)_{B-L}$ gauge interaction to generate such a potential. The effects of such a flavor universal interaction are different from   those of flavor dependent potentials which have been considered before
\cite{wolfenstein,Valle:1987gv,Langacker:1988up,Bergmann:1999rz,Grifols:2003gy,Davidson:2003ha,GonzalezGarcia:2004wg,Barranco:2005ps,GonzalezGarcia:2006vp}, in particular, the effects on long baseline active flavor conversion are small.
In order to obtain a sufficiently large effect to impact short baseline oscillations, we require a new light gauge boson with mass   in the several keV region.

We extend the standard model to include the $B-L$ gauge interaction by making all derivatives covariant under a $B-L$ gauge interaction as well as under the standard model gauge group, and adding the conventional kinetic term for the $B-L$ gauge boson. We refer to this gauge boson as the paraphoton.
Gauge anomaly cancellation is ensured by the addition of 3   neutrinos which are sterile under the standard model interactions and have opposite $B-L$ charge to the usual neutrinos.
As discussed further in \S\ref{chameleonsection}, the paraphoton  acquires a small mass from a condensate of a scalar field $s$. A Yukawa coupling to the condensate will   be responsible for the Majorana masses of the sterile neutrinos.

Because neutrino energies far exceed their mass in oscillation experiments, we can expand the effective Hamiltonian for neutrino propagation:
\be
\mathcal{H} \simeq E\cdot\openone + \frac{\mathcal{M}^{\dag}\mathcal{M}}{2E} + \mathcal{V}
\ee
where $E$ is the neutrino energy, $\mathcal{M}$ is the mass matrix, and $\mathcal{V}$ is the potential matrix. $\mathcal{V}$ receives contributions both from the standard model weak interactions and from the $B-L$ interaction.

Neutrino masses can be Dirac ($B-L$ conserving) or Majorana ($B-L$ violating). Both types of mass terms  are required to give active-sterile mixing.  Using a notation where all fermion fields are 2 component left handed Weyl fields, and suppressing the Lorentz indices, the allowed Dirac mass terms are
\be
m_{ij}\nu_iN_j + \textrm{ h.c.}\ ,
\ee
where we have labeled the sterile neutrinos $N_i$, $i=1,2,3$. These mass terms can arise from a small Yukawa coupling to the usual Higgs boson.
  Majorana masses for the sterile neutrinos  arise from a coupling
    \be
y_{ij}s N_iN_j + \textrm{ h.c.}
\ee
where $y_{ij}$ are dimensionless couplings, and $s$ is a scalar field with $B-L$ charge -2 and no standard model gauge interactions.  Then the sterile neutrino Majorana mass matrix will have entries
\be
M_{ij} = y_{ij}\big<s\big>
\ee
and their mass terms are
\be
M_{ij}N_iN_j + \textrm{ h.c.}
\ee
The flavor basis for sterile neutrinos is irrelevant, so we   take the sterile Majorana mass matrix $\mathcal{M}_s$ to be  a 3 by 3 diagonal matrix, with eigenvalues $M_i$, for $i=1,2,3$.  Therefore, the  mass matrix for all 6 neutrinos  is
\be
\mathcal{M} = \left( \begin{array}{c c}
0 & \mathcal{M}_D \\
\mathcal{M}_D^{T} & \mathcal{M}_s
\end{array} \right)\ ,
\ee
where  $\mathcal{M}_D$ is the 3 by 3  Dirac mass matrix.
Atmospheric and solar oscillation experiments provide good constraints on large mixing angle long baseline active flavor mixing. As we will see, in matter at high energy,  the active neutrinos nearly decouple from the sterile neutrinos, and mainly oscillate between active flavors with parameters governed by the Dirac mass matrix.  Therefore, we will take the mass squared differences  in the Dirac mass matrix to be small.  This makes the Dirac masses almost degenerate, with an overall scale $m$, which we call the Dirac mass from now on.  Then the Dirac mass matrix can be written
\be
\mathcal{M}_D \approx mu
\ee
where $u$ is a unitary 3-by-3 matrix relating the mass and flavor bases.  Long baseline tests are sensitive to small non-degeneracy in the masses, while short baseline tests are insensitive.  Parameterizing $\mathcal{M}_D = mu + \delta m$, with the entries in $\delta m$ much smaller than $m$, we note than  $\delta m$ may be diagonalized in a different basis from $mu$, and that small $\delta m$ has a significant effect on long baseline oscillations but a negligible effect at short baseline. The mixing angles relevant for long baseline depend both on the angles in $u$ and on  $\delta m$, while the parameters relevant for short baseline disappearance and active flavor conversion depend on $u$, $m$ and the three Majorana masses $M_i$. Because   long and short baseline oscillations depend on independent parameters, we do not consider the constraints from long baseline oscillations in our analysis, and neglect   $\delta m$. 

We do not wish to do detailed model building in this paper, but note that  a nonabelian flavor symmetry and a suitable choice of flavor breaking spurions could explain the near degeneracy in the Dirac mass matrix.

Note that, as discussed further in \S\ref{chameleonsection},  the expectation value of $s$  may be affected by the environment \cite{Sawyer:1998ac,Fujii:1990se,mota1,mota2,us,us2,Weiner:2005ac, Hannestad:2005gj,
cham,Vectorcham,Olive:2007aj}, and so the paraphoton mass and neutrino oscillation parameters could be quite different in extreme environments such as stars, supernovae, and the early universe. For this reason the  constraints from astrophysics and cosmology \cite{Grifols:1986fc,Grifols:1988fv,Raffelt:1988rx,Koshiba:1992yb,Grifols:1996fk,Raffelt:1999tx,Raffelt:2000kp,Dolgov:2002wy,Cyburt:2004yc,Hannestad:2005gj,Lesgourgues:2006nd} are complicated and depend on different parameters than the neutrino oscillations. We  assume that the neutrinos masses are  constant in our analysis of terrestrial experiments, although it could be interesting to consider environmental dependence  of the neutrino masses \cite{Sawyer:1998ac,us2,Zurek:2004vd}.

We can now write the mass and potential matrices in the flavor basis in terms of 3-by-3 blocks:
\be
\mathcal{M} = \left(
\begin{array}{c c}
0 & mu \\
mu^{T} & \mathcal{M}_s
\end{array} \right)
\qquad \mathcal{V} = \left(
\begin{array}{c c}
-V\cdot\openone & 0 \\
0 & V\cdot\openone
\end{array} \right)
\ee
We rotate from the flavor basis to a basis where the Dirac mass matrix $mu$ in $\mathcal{M}$ is diagonal, via the unitary matrix $\mathcal{U}$, which leaves $\mathcal{V}$ unchanged:
\be
\begin{split}
\mathcal{U} &= \left(
\begin{array}{c c}
u & 0 \\
0 & \openone
\end{array} \right) \\
\mathcal{M}' = \mathcal{U}^T\mathcal{M}\mathcal{U} &= \left(
\begin{array}{c c}
0 & m\openone \\
m\openone & \mathcal{M}_s
\end{array} \right) \qquad \mathcal{V}' = \mathcal{V}
\end{split}
\ee
We call this the prime basis. In this basis the effective Hamiltonian for neutrino evolution is,
\be\label{hprime}
\mathcal{H}' = E + \frac{m^2}{2E} + \frac{1}{2E} \left(
\begin{array}{c c}
-2VE\cdot\openone & m\mathcal{M}_s \\
m\mathcal{M}_s & 2VE\cdot\openone + \mathcal{M}_s^2
\end{array} \right)\ .
\ee
Note that we are neglecting the usual MSW terms from the weak interactions, as these have negligible effect at short baseline.
The Hamiltonian in the prime basis is 2-by-2 block diagonal, with each block having the form
\be
\mathcal{H}'_2 = E + \frac{m^2}{2E} - V + \frac{1}{2E} \left(
\begin{array}{c c}
0 & mM_i \\
mM_i & 4VE + M_i^2
\end{array} \right)\ .
\ee
 
If the sign of $V$ is positive, then provided that $m \ll M_i$, we have  a mini seesaw with small active-sterile mixing.  In this case the lighter neutrinos are almost entirely active, while the heavier are almost all sterile.  Furthermore, the probability of active neutrino disappearance into sterile is suppressed by the ratio $m^2/M_i^2$.  As the energy of the experiment increases, the active-sterile mixing  is suppressed further.  This high energy limit is relevant for MiniBooNE, which has an energy that is an order of magnitude larger than LSND, and can therefore have highly suppressed oscillations.  Note that the small mass differences in the mostly active neutrinos contribute essentially nothing to oscillations at MiniBooNE and LSND, just as the standard oscillation model predicts no oscillations at these experiments, but will be important when considering long baseline oscillation experiments.

If the sign of $V$ is negative, as is the case for anti-neutrinos, then there can be a range of resonance energies for which $4VE \sim M_i^2$, giving  large mixing between the active and sterile sectors.    This enhances oscillations between active flavors, and may account for oscillations seen in anti-neutrinos at LSND. Oscillations are suppressed when $|4 VE + M_i^2|\gg mM_i$, so the effects of sterile neutrinos on   anti-neutrino  oscillation experiments should be small at high energy, but larger than on neutrino oscillations.

\section{Oscillation Probabilities}

Here we find the formula for oscillation between active flavors.  We pay special attention to those experiments where the potential could be relevant, and discuss the implications of solar and atmospheric neutrino experiments.  We outline the constraints on the B-L gauge interaction parameters from electroweak measurements and discuss other possible constraints.  Finally, we fit to the results of LSND and MiniBooNE, and make a prediction for anti-neutrino oscillations at MiniBooNE.

To find the oscillation probability, we diagonalize the Hamiltonian and find the transformation between the flavor basis and diagonal basis.  In the prime basis, we utilize the block diagonal nature of the Hamiltonian.  Each 2-by-2 block in (\ref{hprime}) is diagonalized via the unitary angle $\theta_i$, where
\be
\tan 2\theta_i = \frac{2mM_i}{4VE + M_i^2}
\ee
These angles define the unitary matrix needed to diagonalize the Hamiltonian in the prime basis:
\be
\begin{split}
\mathcal{H}_D &= \mathcal{U}'^T\mathcal{H}'\mathcal{U}' \\
&= E + \frac{m^2}{2E} - V + \frac{1}{2E}\textrm{diag}(\lambda_1^-,\lambda_2^-,\lambda_3^-,\lambda_1^+,\lambda_2^+,\lambda_3^+)
\end{split}
\ee
with
\be
\mathcal{U}' = \left(
\begin{array}{c c c c c c}
\cos\theta_1 & 0 & 0 & \sin\theta_1 & 0 & 0 \\
0 & \cos\theta_2 & 0 & 0 & \sin\theta_2 & 0 \\
0 & 0 & \cos\theta_3 & 0 & 0 & \sin\theta_3 \\
-\sin\theta_1 & 0 & 0 & \cos\theta_1 & 0 & 0 \\
0 & -\sin\theta_2 & 0 & 0 & \cos\theta_2 & 0 \\
0 & 0 & -\sin\theta_3 & 0 & 0 & \cos\theta_3 \\
\end{array} \right)
\ee
and
\be
\begin{split}
\lambda_i^{\pm} &= \frac12\Big(4VE + M_i^2 \pm \sqrt{4m^2M_i^2 + (4VE + M_i^2)^2}\Big) \\
&= \pm m^2\frac{M_i}{m}\big(\tan\theta_i\big)^{\mp1} \\
\cos^2\theta_i &= \frac12\bigg(1 + \frac{4VE + M_i^2}{\sqrt{4m^2M_i^2 + (4VE + M_i^2)^2}}\bigg)
\end{split}
\ee
Between the flavor basis and the diagonal basis, we have made the transformation
\be
\mathcal{H}_D = \mathcal{U}^T\mathcal{U}'^T\mathcal{H}\mathcal{U}'\mathcal{U} = \mathcal{U}_D^T\mathcal{H}\mathcal{U}_D
\ee
The oscillation probability for neutrinos of a specific energy $E$ over a baseline $L$ is given by the standard oscillation formula \cite{PDbook},
\be
P_{a\rightarrow b} (L,E) = \sum_{\alpha,\beta} \Big|K_{ab,\alpha\beta}\Big|\cos\Big(\tilde{\Delta}_{\alpha\beta}L - \textrm{arg}\big(K_{ab,\alpha\beta}\big)\Big)
\ee
where
\be
\begin{split}
K_{ab,\alpha\beta} &= \mathcal{U}_{D}^{a\alpha}\textrm{ }\mathcal{U}_{D}^{b\alpha}{}^*\textrm{ }\mathcal{U}_{D}^{a\beta}{}^*\textrm{ }\mathcal{U}_{D}^{b\beta} \\
\tilde{\Delta}_{\alpha\beta} &= \Big|\big(\mathcal{H}_D\big)_{\alpha\alpha} - \big(\mathcal{H}_D\big)_{\beta\beta}\Big|
\end{split}
\ee
 
For simplicity in the numerical analysis we will use  the values $\theta_{13} = 0$ and $\theta_{23} = \pi/4$, the angles   suggested by long baseline oscillation experiments, although in principle, the angles need not be related. The probability of oscillation between electron and muon neutrinos is
\be
\begin{split}
P_{\mu\rightarrow e}(L,E) = 2s_{12}^2c_{12}^2\Big[&\cos^2\theta_1\cos^2\theta_2\sin^2\Big(\frac{L}{2E}\big(\lambda_1^- - \lambda_2^-\big)\Big) + \sin^2\theta_1\sin^2\theta_2\sin^2\Big(\frac{L}{2E}\big(\lambda_1^+ - \lambda_2^+\big)\Big) \\
&+ \cos^2\theta_1\sin^2\theta_2\sin^2\Big(\frac{L}{2E}\big(\lambda_1^- - \lambda_2^+\big)\Big) + \sin^2\theta_1\cos^2\theta_2\sin^2\Big(\frac{L}{2E}\big(\lambda_1^+ - \lambda_2^-\big)\Big) \\
&- \cos^2\theta_1\sin^2\theta_1\sin^2\Big(\frac{L}{2E}\big(\lambda_1^- - \lambda_1^+\big)\Big) - \cos^2\theta_2\sin^2\theta_2\sin^2\Big(\frac{L}{2E}\big(\lambda_2^- - \lambda_2^+\big)\Big)\Big]
\end{split}
\ee
Before moving on to numerical analysis, we discuss constraints on the parameters governing the oscillation probability from experiments.  Long baseline oscillation tests help constrain the parameters affecting the oscillation probability.  Experiments measuring neutrino disappearance are of interest, as they constrain the ratio of the Dirac and Majorana masses.  Additionally, we discuss electroweak constraints on $U(1)_{\textrm{\tiny{B-L}}}$, and show there is sufficient parameter space for a strong potential.

We concentrate our analysis on fitting the positive anti-neutrino oscillation results of LSND and the null oscillation results of MiniBooNE above 475 MeV.  We then discuss the speculative low energy excess seen at MiniBooNE and the ability of the model to accommodate this result.  Finally, we give a prediction for anti-neutrino oscillations at MiniBooNE.

\section{Constraints on Parameters from Neutrino Oscillation Experiment}

Before undertaking the numerical analysis, we consider constraints on parameters for the oscillation probability from experiment.  The parameter space is large,  and we make some simplifying assumptions.

If the eigenvalues of the Dirac matrix are exactly equal, then in the small $m/M$ limit,  the matrix $u$ will be the matrix   governing mixing in the active sector as given by solar and atmospheric oscillation experiments in (\ref{stdangles}). We make this assumption, and  pick  the most straightforward values for $\theta_{13}$ and $\theta_{23}$ suggested by these experiments, $\theta_{13} = 0$ and $\theta_{23} = \pi/4$.  We use the central value for the remaining angle, $\tan^2\theta_{12} = 0.45$.

The Dirac and Majorana neutrino masses $m$ and $M_i$ are loosely bound by experiments.  The Dirac neutrino mass sets the scale for the Majorana neutrino masses, since their relative size governs the mixing between the active and sterile sectors.  We take the Dirac mass to be sufficiently small, while still maintaining degeneracy in the neutrino masses, in the range $m = 0.1 - 0.4$ eV.  Tests for active neutrino disappearance will constrain the Majorana masses $M_i$ based on the range for $m$.  In general, the Majorana mass will be in the range 1 - 10 eV.

Electron neutrino disappearance tests give the most stringent constraint on  $M_i$ relative to $m$, as does the measurement of the total active solar neutrino flux at SNO \cite{Ahmed:2003kj}.  Muon neutrino disappearance and conversion  experiments \cite{Dydak:1983zq,Stockdale:1984ce,Fukuda:2000np,Ambrosio:2001je,Ahn:2002up,
Michael:2006rx} involve baselines through matter at higher energies  where the potential term suppresses the effect of the sterile neutrinos, so they will not provide a constraint. The best electron neutrino disappearance constraint comes from the CHOOZ experiment \cite{Apollonio:1999ae,Apollonio:2002gd}.  Anti-electron neutrinos at CHOOZ have an energy of a few MeV, and propagate approximately 1 km through air before detection.  The low B-L charge density in air coupled with the low neutrino energy means the potential term has essentially no effect.  The SNO  results are less sensitive to the disappearance of electron neutrinos   than is the CHOOZ experiment.

Both experiments are sensitive to oscillations with neutrino mass squared differences near $\Delta m_{sun}^2$, far below the scales $m$ and $M_i$ relevant for oscillations between active and sterile neutrinos.  Therefore, the oscillatory terms average out, and we can replace the oscillatory $\sin^2$ terms with $\frac12$.  The rate of electron neutrino disappearance into sterile neutrinos is therefore given by
\be
P(\nu_e\rightarrow\nu_s) = 2c_{12}^2\cos^2\theta_1\sin^2\theta_1 + 2s_{12}^2\cos^2\theta_2\sin^2\theta_2
\ee
where
\be
\sin^2\theta_i \approx \frac{m^2}{M_i^2} + \mathcal{O}\Big(\frac{m^4}{M_i^4}\Big)
\ee
SNO provides a measure of the total neutrino flux that agrees well with solar models, with a 9\% experimental error and 16\% theoretical error in the solar model \cite{Bahcall:2004fg,Bahcall:2005fg}.  CHOOZ provides a better constraint in this case, having found no evidence for $\bar{\nu}_e$ disappearance with a 3.9\% total uncertainty.  This gives a constraint
\be\label{CHOOZ}
0.039 > 1.37\frac{m^2}{M_1^2} + 0.62\frac{m^2}{M_2^2}
\ee
The interference between oscillations involving the different eigenstates should average to zero, unless the masses are extremely degenerate.  We  have separate absolute bounds on $M_1$ and $M_2$:
\be
\frac{M_1}{m} > 6 \qquad \qquad \frac{M_2}{m} > 4 \ .
\ee
Since the sterile and active sectors decouple as $M_i/m \rightarrow \infty$, the only upper bound on this ratio will be set by LSND and MiniBooNE, when the sterile sector becomes relevant.

The scale for the potential $V$ is set by a number of considerations.  Since it becomes relevant for LSND and MiniBooNE, in range of MiniBooNE energies the potential term $4VE$ should be comparable to or larger than $M_i^2$.  Specifically, for oscillations at higher energies at MiniBooNE to be suppressed, we require the limit
\be
V>\frac{M_i^2}{4 E_{\textrm{\tiny{MB}}} } \ .
\ee
At LSND, however, oscillations should not be highly suppressed, at least not for anti-neutrinos. Therefore  the potential should not be large compared to the quantity  $M_i^2/(4 E_{\textrm{\tiny{LSND}}})$. Since $E_{\textrm{\tiny{LSND}}} \sim 0.1E_{\textrm{\tiny{MB}}}$, this implies
\be
\frac{M_i^2}{4 E_{\textrm{\tiny{MB}}}}  \lsim V \lsim  \frac{M_i^2}{4 E_{\textrm{\tiny{LSND}}}}  \ .\ee
Thus the  optimal range for $V$ is determined by the neutrino energy at the LSND and MiniBooNE experiments and the range of Majorana masses $M_i$.  We cannot have $VE_{\textrm{\tiny{LSND}}}$ too large, and the range of $M_i^2$ should be consistent with the LSND analysis and KARMEN constraint \cite{Armbruster:2002mp}, in the eV range. Therefore, we have bounds for the parameters that control the oscillation probability at LSND and MiniBooNE.  After describing constraints on the vector mass and gauge coupling of $U(1)_{\textrm{\tiny{B-L}}}$, which constrain the potential $V$, we go on to describe our numerical analysis.

\subsection*{Constraints on Weakly Coupled $U(1)_\textrm{\tiny{{B-L}}}$}
Precision electroweak measurements and rare decays constrain the coupling and vector mass for a $U(1)$ gauge interaction \cite{Fayet:1980rr,Nelson:1989fx,Fayet:1990wx,Leike:1998wr,Aranda:2000ma,Chang:2000xy,Fayet:2006sp,Badertscher:2006fm,Fayet:2007ua}.  Collider experiments can only constrain $B-L$ if the  coupling is relatively large, and the parameters we consider here are far away from this region.  The most stringent constraint arises from corrections to $g-2$ of the electron and muon.  To leading order in the coupling, the electron $g-2$ correction goes as
\be
\delta a_e \approx \left\{
\begin{array}{c}
\displaystyle\frac{  g^2}{12\pi^2}\frac{m_e^2}{m_v^2} \qquad m_v \gg m_e \\
\\
\displaystyle\frac{  g^2}{8\pi^2} \qquad \quad m_v \ll m_e
\end{array} \right.
\ee
where $m_v$ is the paraphoton  mass and $g$ is the gauge coupling.  The correction to the muon is the same, but with $m_e \rightarrow m_{\mu}$.  The experimental bounds on corrections to the electron and muon $g-2$ are \cite{Gabrielse:2006gg,Bennett:2006fi}
\be
\begin{split}
\delta a_e^{\textrm{\tiny{exp}}} &= 1.5\cdot10^{-12} \\
\delta a_{\mu}^{\textrm{\tiny{exp}}} &= 1.2\cdot10^{-9}
\end{split}
\ee
The potential neutrinos experience in matter is generated by low momentum transfers via the $B-L$ photon between charges in matter and neutrinos.  The strength of the potential is
\be
V = \frac{g^2}{m_v^2}\rho_{\textrm{\tiny{B-L}}}
\ee
where $\rho_{\textrm{\tiny{B-L}}}$ is the $B-L$ charge density in matter, typically  about $10^{30}$ charges/m$^3$ for rock at the Earth's surface.  Since we want potentials stronger than 0.1 neV, we have a bound on $g/m_v$:
\be
\frac{g}{m_v} > 1.14\cdot10^{-10} \textrm{ eV}^{-1} \Rightarrow \frac{g}{10^{-5}} > \frac{m_v}{88 \textrm{ keV}}
\ee
Since $g$ must be small to meet the bounds from the electron and muon $g-2$, we need $m_v < m_e$.  Then the best constraint comes from the electron $g-2$, which, when combined with the independent  measurement of $\alpha$ from the quantum Hall effect \cite{PDbook}, bounds $g$:
\be
g <  1.7 \cdot10^{-5}
\ee
To have potentials stronger than 0.1 neV, with $g$  satisfying this bound, constrains $m_v$: $m_v <$ 150 keV.  If we use the lower limit of the potential found in numerical analysis, $V  > 3$ neV, and thus the vector mass must be lighter than 28 keV.

For such a weak coupling, constraints on the new boson from rare particle decays  are well within the bounds.  The rate for any process with a paraphoton replacing a photon, such as $e^+e^-\rightarrow\gamma\gamma^*$, is suppressed by a factor of $g^2/e^2$, which from the limit above, is
\be
\frac{g^2}{e^2} < 3\cdot10^{-9}
\ee
which is   below the measured limits on rates of rare processes.
\section{Astrophysics}
Note that the interactions of light gauge bosons with quarks, leptons and neutrinos, as well as the oscillations of sterile neutrinos,  are severely constrained from their contributions to energy loss and energy transfer in stars and supernovae \cite{Grifols:1988fv,Raffelt:1999tx}.  We can avoid the stellar constraints if the effective paraphoton mass inside the star is  much heavier than the temperature.  For constant mass paraphotons these constraints pose a challenge to obtaining a sufficiently large potential to fit all the neutrino oscillation data. However the environment in the core of a  star or supernova   is  extreme, with   densities of $B-L$ and neutrinos much higher than in the earth.  
In the next section we show that in a dense large object the effective mass of the paraphoton can be larger than it is on the earth, increasing the parameter space for paraphotons.
 
\subsection{Chameleon effect in stars}
\label{chameleonsection}
Note that we have assumed that the paraphoton mass is due to the Higgs mechanism from a scalar $s$ with $B-L$ charge of 2. The field $s$ is neutral under the standard model gauge group. The terms in the Lagrangian involving this scalar  and the paraphoton field $B_\mu$ are
  
  \be\CL\supset  (\partial_\mu-2 i g B_{\mu}) s^*(\partial^\mu+2 ig B^\mu) s  - \frac{\lambda_s}{2} |s|^4 +m_s^2 |s|^2\ldots \ee

Note that $s$ has no direct coupling to ordinary   matter  or the active neutrinos.
 However ordinary matter has a net $B-L$ charge density $\rbml$, which acts as a source for the $B_0$ field. The $B_0$ field in turn acts as  negative term in the potential  for $s$, increasing its expectation value and the effective mass of the paraphoton.  Of course the $B_0$ field is not gauge invariant, and neither is the  phase of the $s$ field.  However the screening length, or inverse mass of the paraphoton, is a gauge invariant physical quantity which we wish to determine.
 
 To determine the paraphoton mass inside  large objects,
we    consider  a  time independent spatially constant background charge $B-L$ charge density $\rbml$.     The   $s$ field carries a charge density $\rho_s$ which cancels the background charge.
\be \rho_s= -i 2( D_0 s s^*-s D_0 s^*)\ .
\ee
For a minimal energy configuration of given charge, we take $s$ to have the form $|s| e^{-i 2 w t}$.
Since the phase of $s$ is not gauge invariant
  it is convenient to define a gauge invariant field \be V\equiv w +   g B_0\ .\ee
The charge density  carried by the $s$ condensate is $-4V |s|^2$ while   $V$ is the condensate energy per unit charge, which is  identical to the neutrino potential energy used in the previous sections.
   
      To minimize the total energy for a static configuration where  $V$ and $|s|$ are spatially constant, one must solve  the equations
  \beq
\label{eom}
 \lambda_s |s|^2&= &m^2+  4V^2 \cr
\rho  &=&  4V |s|^2\ .
\eeq
  The solution is
\be\label{wsol2}
V=\frac{ m^2 2^{\frac43} -3\left(\sqrt{(\lambda_s\rho)^2+\frac{16m^6}{27} }-\lambda_s\rho\right)^{\frac23} }{  2^{\frac23}6\left(\sqrt{(\lambda_s\rho)^2+\frac{16m^6}{27} }-\lambda_s\rho\right)^{\frac13}}\ .\ee
and
\be\label{ssol}
|s|=\sqrt{\frac{\rho}{4V}}\ .\ee

In the limit 
\be\label{condition1} |m|  \ll (\lambda\rho)^{\frac13}\ee 
an approximate solution is
\beq
V&\approx&\left(\frac{(\lambda_s\rho)}{16}\right)^{\frac13}\cr
|s|&\approx& \left(\frac{ \rho}{2\sqrt{\lambda_s} }\right)^{\frac13} \  . 
\eeq

 Note that the effective paraphoton mass is $\sqrt{8} g |s| $.

 The chameleon effect is   significant when the value of $\lambda$ satisfies   \Eref{condition1}. The constraint in turn bounds the mass of the physical scalar excitations of the condensate.  
 
 In the opposite limit
 \be\label{condition2} |m|  \gg (\lambda_s\rho)^{\frac13}\ee 
the chameleon effect is negligible and an approximate solution is
\beq
V&\approx&\frac{(\lambda_s\rho)}{4 m^2}\cr
|s|&\approx& m/\sqrt{\lambda_s}\  . 
\eeq

 \subsection{Numerical estimate of chameleonic paraphoton effect on stellar energy loss}
 \label{stellar}
 In this section we give a numerical example demonstrating how the chameleon effect can suppress the energy loss in stars to paraphotons. This is important because constraints from stellar evolution require that at densities between  $0.6 \times 10^4 $g/cm$^3$ and temperatures around $0.7\times 10^8$K,   and also for higher density and temperature  of $2 \times 10^5$g/cm$^3$, and 
  $10^8$ K, radiation of exotic particles should not lead to energy loss greater than about 10 ergs/(g-s)\cite{Yao:2006px}.
 
For simplicity in the   analysis of stellar constraints we take   $g=10^{-5}, \ m_v=28$ keV. In order to have a significant chameleon effect at stellar densities  we take a very flat potential for the scalar field,  $m_s= 10^{-8}$ eV, and $\lambda_s=  5\times10^{-35}$.  (Note that  because ordinary matter does not act as a source for the scalar field, such a light   $s$ field is consistent with constraints from searches for new forces.)  The paraphoton mass at typical earth densities of $2 g/$cm$^3$ is then nearly the same as in vacuum, while at a density of  $0.6 \times 10^4 $g/cm$^3$, assuming about 1/2 of the baryons are neutrons,  the paraphoton mass is 260 keV. A  typical temperature for stellar cores of such a density is $0.7 \times 10^8$ K and so the Boltzmann suppression factor for paraphoton emission is about $10^{-19}$. The emission rate for paraphotons would then be similar to the rate for emission of light vector bosons with an  $\alpha $ of order $10^{-30}$, which is smaller than the quoted stellar evolution limit \cite{Grifols:1988fv} for that density of $  10^{-27}$. At a density of  $2 \times 10^5$g/cm$^3$ the paraphoton mass is 800 keV, and the Boltzmann suppression factor at a temperature of $10^8$ K  is completely negligible.

\subsection{Supernova 1987A}

Weak interactions of bosons lighter than $\sim100 $ MeV are constrained by the observation of a small number of neutrino events lasting for about 10 seconds after the explosion of SN 1987A \cite{Bionta:1987qt,Hirata:1987hu,Yao:2006px}.   If   the paraphoton is much lighter than the supernova temperature, couplings stronger than about $10^{-2}$ are excluded, as neutrinos would then remain thermally coupled until lower temperatures than suggested by the energies detected \cite{Fayet:2006sa}, while couplings weaker than about $10^{-6}$ are also excluded, as then the supernova would cool too rapidly by emitting paraphotons \cite{Hoffmann:1987et,Raffelt:2000kp}.

For the parameters considered in \S\ref{stellar}, at a typical nuclear density of $10^{14}$g/cm$^3$ found in a supernova core,  the paraphoton mass would be of order  700 GeV.  Even if the potential is not this flat, and so the paraphoton is  lighter than the temperature and in thermal equilibrium with the neutrinos, it is strongly enough coupled to not lead to excessively rapid supernova cooling . 

One also must consider the question of neutrino emission. Usually the supernova observation provides severe constraints on sterile neutrinos because they are more weakly coupled and can more easily escape than neutrinos.   The sterile neutrino interaction cross section depends on the paraphoton mass.   If the paraphoton is light the sterile neutrino cross section from paraphoton exchange at a temperature below 100 MeV is larger than the weak cross section, and so sterile neutrinos would not escape more easily than ordinary neutrinos. For a paraphoton mass of 700 GeV the sterile neutrino productions cross section is of order $10^{-28}$ times the weak cross section and so production of sterile neutrinos may be neglected.
A range of intermediate high density paraphoton masses will be excluded by supernova cooling, however astrophysical constraints are not the main focus of this paper and so this range will not be computed here. 

One also must consider sterile neutrino production via neutrino oscillations. However,  for such large values of the B-L potential as found in supernova core, active sterile neutrino mixing angles are negligible.

\subsection{Nucleosynthesis constraints}

The mechanism used for baryogenesis will determine the net $B-L$ charge of the universe, and the $B-L$ charge density in the early universe. Electroweak baryogenesis, for instance, gives a net $B-L$ charge of zero \cite{Kuzmin:1985mm},  while Affleck-Dine baryogenesis \cite{Affleck:1984fy} could give a large net $B-L$ density. 

If we assume a net $B-L$ charge of zero, then the main change to the effective potential for $s$ comes from thermal corrections, including the neutrino mass contribution to the energy density. Even with a potential as flat as assumed in section \S\ref{stellar}, these only increase  the $s$ vev by a factor of about a 100, and so the paraphoton and the sterile neutrinos are   lighter than the temperature during nucleosynthesis. The paraphoton may be produced singly and will be in thermal equilibrium, acting approximately as another neutrino species. The cross section for sterile neutrino interactions at nucleosynthesis temperatures from paraphoton  is comparable to weak, and so the total number of effective neutrino species during nucleosynthesis would be 7, a value inconsistent with the Helium abundance \cite{Yao:2006px}.

Another attractive  mechanism, available in supersymmetric theories, is Affleck-Dine baryogenesis, which relies on large scalar field vevs in the early universe along flat directions of the scalar potential. This mechanism could   create a large net $B-L$ density along a flat direction where $s$ has a large vev. The minimum of the effective potential for $s$ would then decrease with the Hubble expansion. This has three potential effects on nucleosynthesis. First, a large lepton number chemical potential would affect the Helium abundance, and allow for more light relativistic species \cite{Barger:2003rt,Kneller:2004jz} Secondly, even it the lepton number chemical potential is not large compared with the nucleosynthesis temperature, the paraphoton    and sterile neutrino masses  could always be larger than the temperature, and so these are never in equilibrium. Third, even if the sterile neutrinos are lighter than the temperature, a heavier paraphoton reduces the cross section for sterile neutrino production, and so they need  not be in thermal equilibrium.
Note that sterile neutrino production via neutrino oscillations will be negligible due to tiny active-sterile mixing angles at high $B-L$ charge density and large $s$ vev.

\section{Numerical Analysis}
We do a numerical analysis of the oscillation probability at LSND and MiniBooNE to find the region of parameter space that fits both of these experiments.  We scan over parameters that satisfy the CHOOZ electron disappearance bounds. For these parameters the bounds on muon neutrino short baseline disappearance \cite{Dydak:1983zq} are also satisfied. We then make a quantitative prediction for the results of MiniBooNE anti-neutrino oscillations.  We are interested in fitting the anti-neutrino oscillation results of LSND and the null oscillation results above low energy at MiniBooNE.  The   low energy excess seen at MiniBooNE and   anti-neutrino oscillations at MiniBooNE may then be considered in the context of these two results.  The results can be summarized in terms of the oscillation probability:
\be\label{constraints}
\begin{split}
&P_{\textrm{\tiny{LSND}}}(\bar{\nu}_{\mu}\rightarrow\bar{\nu}_e) = (0.264\pm0.081)\% \\
&P_{\textrm{\tiny{MB}}}(\nu_{\mu}\rightarrow\nu_e,\textrm{ }475\textrm{ MeV} < E < 3 \textrm{ GeV}) < 0.064\% \\
\end{split}
\ee
For each experiment, the energy distribution of neutrinos must be taken into account.  We use the Michel spectrum from the muon decay giving rise to $\bar{\nu}_{\mu}$ at LSND, and fit an interpolating function to the approximate $\nu_{\mu}$ energy spectrum at MiniBooNE \cite{Tanaka:2007yk}.  If $s(E)$ is the relevant energy density function, then the oscillation probability at a baseline $L$ is just
\be
P(L) = \int dEs(E)P(L,E)
\ee
and we do not consider detector geometry at either experiment.

We pick the following values for the parameters, and study the regions listed below for the remaining ones:
\be\label{anglespace}
\begin{split}
\theta_{12} = 0.591 \qquad \theta_{13}& = 0 \qquad \theta_{23} = \pi/4 \\
0.1 \textrm{ neV} < V < 5 \textrm{ neV} \quad & \quad 0.1 < m < 0.4 \textrm{ eV} \\
6m < M_1 < 7 \textrm{ eV} \quad & \quad 4m < M_2 < 7 \textrm{ eV} \\
0.039 > 1.37&\frac{m^2}{M_1^2} + 0.62\frac{m^2}{M_2^2}
\end{split}
\ee
The last limit above imposes the CHOOZ constraint in ref. \cite{Apollonio:2002gd}.  We choose an upper bound on $V$ to place a reasonable lower bound (10 keV) on the paraphoton mass $m_v$.  The lower limit on $V$ and the upper limit on $M_i$ cut off the analysis region where changes in the magnitude of those variables have a small effect on the oscillation probability.  As we will see, those limits do not serve as real constraints on the parameter space, with very few points saturating the limits.

\subsection*{Results}
For the analysis, we sample the parameter space with  $3\cdot10^6$ points.  At each point, we require the LSND anti-neutrino result and the MiniBooNE null result to be satisfied, as well as the CHOOZ constraint.  For the points that pass the constraints, we find both the MiniBooNE low energy oscillation probability and the MiniBooNE anti-neutrino oscillation probabilities at high and low energy.   The total volume of the parameter space we fit is 9.95 eV$^3\cdot$ 4.9 neV, giving a mean volume spacing between points of (0.05 eV)$^3\cdot$ 0.11 neV.  Because the oscillation probability is smooth, we are confident that this coverage is sufficient to find interesting regions in the parameter space.

Approximately 0.1\% of the trial points fit LSND and MiniBooNE and pass the CHOOZ constraint, meaning there is an ample region of parameter space which reconcile LSND and MiniBooNE.  Of the points fitting the constraints, the parameters fall within the ranges:
\be
\begin{split}
2.5 \textrm{ neV} < V < 5 \textrm{ neV} \quad & \quad 0.2 \textrm{ eV} < m < 0.35 \textrm{ eV} \\
2.5 \textrm{ eV} < M_1 < 7 \textrm{ eV} \quad & \quad 0.9 \textrm{ eV} < M_2 < 1.7 \textrm{ eV}
\end{split}
\ee
Many points sit near the CHOOZ bound, implying it provides a real constraint on the model.  Conversely, the lower limits for $m$ and $V$, and the upper limit on the Majorana masses $M_i$ that we have chosen do not seriously constrain the ability of the model to reconcile LSND and MiniBooNE, as few points in the fits lie close to these bounds.  Therefore, we conclude that the parameter space we have chosen is suitable to test whether  the model can reconcile LSND and MiniBooNE.

\subsection*{The Low Energy Excess at MiniBooNE}
For the points passing the fits, we find the model is capable of fitting a portion of the observed low energy excess at MiniBooNE.  If the low energy excess observed at MiniBooNE is determined to be a signal of $\nu_{\mu}\rightarrow\nu_e$ transmutation, then we are obviously interested in fitting such a dramatic excess with this model.  However, we find difficulty fitting the entire excess while maintaining  the observed low values of the oscillation probability at LSND and MiniBooNE (above 475 MeV).  In the energy range of 200 to 475 MeV, we fit at most approximately 40\% of the observed excess, corresponding to an oscillation probability of 0.5\%.  A number of points have a small low energy excess, which are suitable to reconcile LSND and MiniBooNE if the observed MiniBooNE excess reduces with further analysis. 

We have restricted the parameter space (see eq. \ref{anglespace})  in order to make the numerical analysis faster.   In particular we do not vary the mixing angles in the matrix $u$, which control the  flavor fractions of the active component of the heavier neutrinos. In our analysis, only 2 of the heavy neutrinos have any electron flavor component.   One effect of varying the  angles  in $u$ would be to allow all three sterile neutrinos to play a role.  Allowing one of the heavy  neutrinos to have a small electron flavor component would allow it play a role in electron flavor conversion while being lighter than $4 m$ without violating the CHOOZ constraint. Should the low energy MiniBooNE electron excess survive further  analysis, it would be interesting to see if the model can give  the entire excess when more parameters are varied. However we believe we have varied enough parameters to cover the essential implications of the model.

In Figure 1, we give a histogram of the low energy excess for the fit points, and in Figure 2 we show the correlations between the low energy excess and the oscillation probability above 475 MeV at MiniBooNE.  Figure 2 shows the main difficulty in fitting the low energy excess  in the model --- there is a   correlation between the well constrained MiniBooNE oscillation probability above 475 MeV and the low energy excess.  

$\\$
\begin{figure}[h!]
\begin{center}
\includegraphics[width=8.0cm]{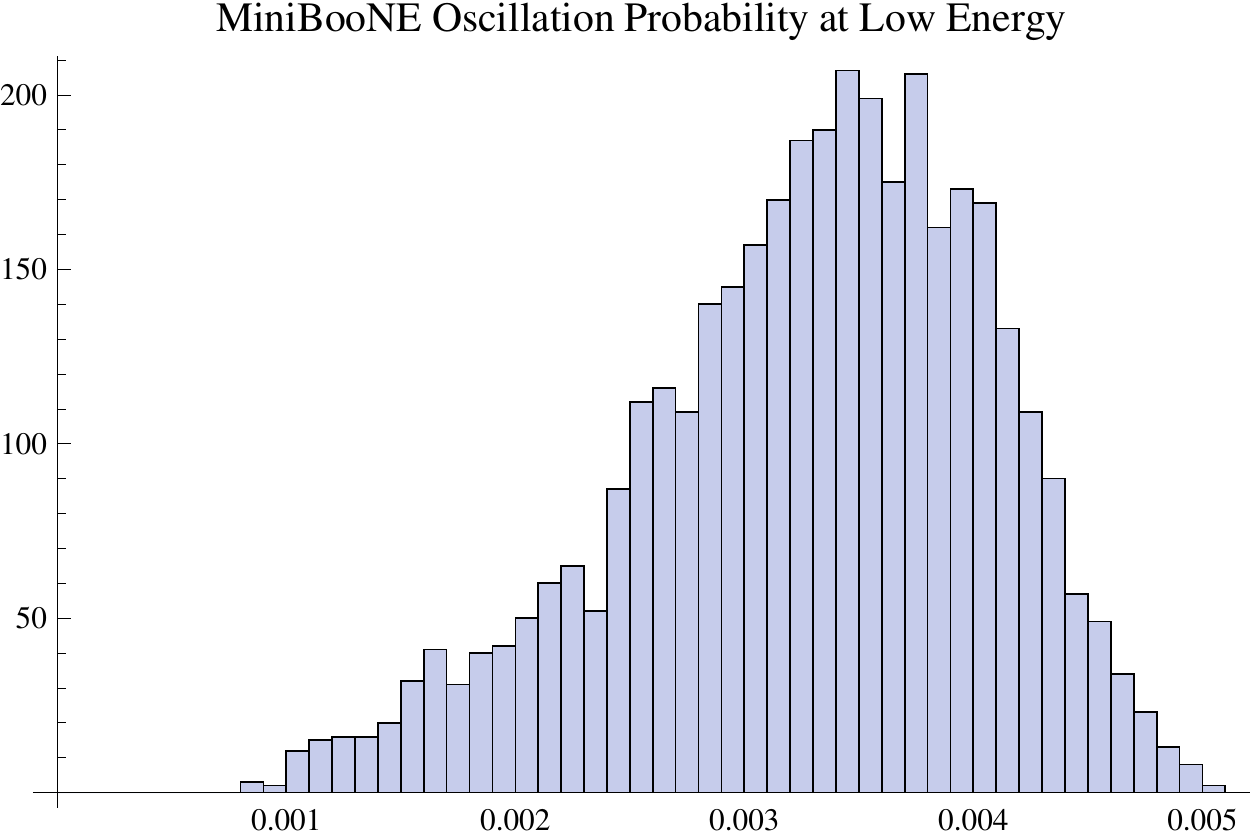}
\caption{A histogram of the MiniBooNE oscillation probability for neutrinos in the energy region from 200 to 475 MeV}
\label{MBloE}
\end{center}
\end{figure}

\begin{figure}[h!]
\begin{center}
\includegraphics[width=8.0cm]{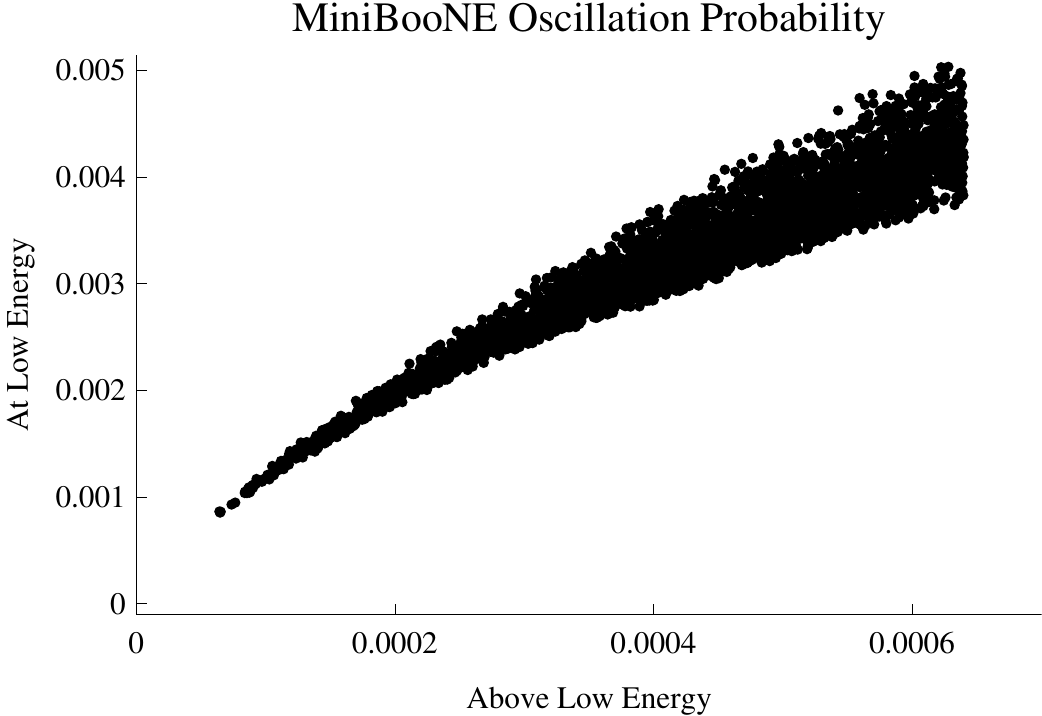}
\caption{The average  MiniBooNE neutrino oscillation probability in the  energy region  from 200- 475 MeV vs. the region above  475 MeV}
\label{MBtoMBloE}
\end{center}
\end{figure}

\subsection*{Predictions for Anti-Neutrino Oscillations at MiniBooNE}
We can make definite predictions for anti-neutrino oscillations at MiniBooNE.  Because $\big|VE_{\textrm{\tiny{MB}}}\big|$ must be similar to the Majorana masses $M_i^2$ to reconcile LSND and MiniBooNE, in anti-neutrinos we expect an MSW type  resonance near or in the low energy MiniBooNE region which leads to relatively large $\bar{\nu}_{\mu}\rightarrow\bar{\nu}_e$ conversion.  At our fit points, we can examine the oscillation probability in anti-neutrinos and correlate the predicted signal with the oscillation probability in neutrinos at MiniBooNE.  In our analysis, we assume that the energy spectrum of anti-neutrinos is the same as neutrinos at MiniBooNE.     In Figure 3, we give the anti-neutrino oscillation probability in the low energy region from 200- 475 MeV versus the region above  475 MeV.

$\\$
\begin{figure}[h!]
\begin{center}
\includegraphics[width=8.0cm]{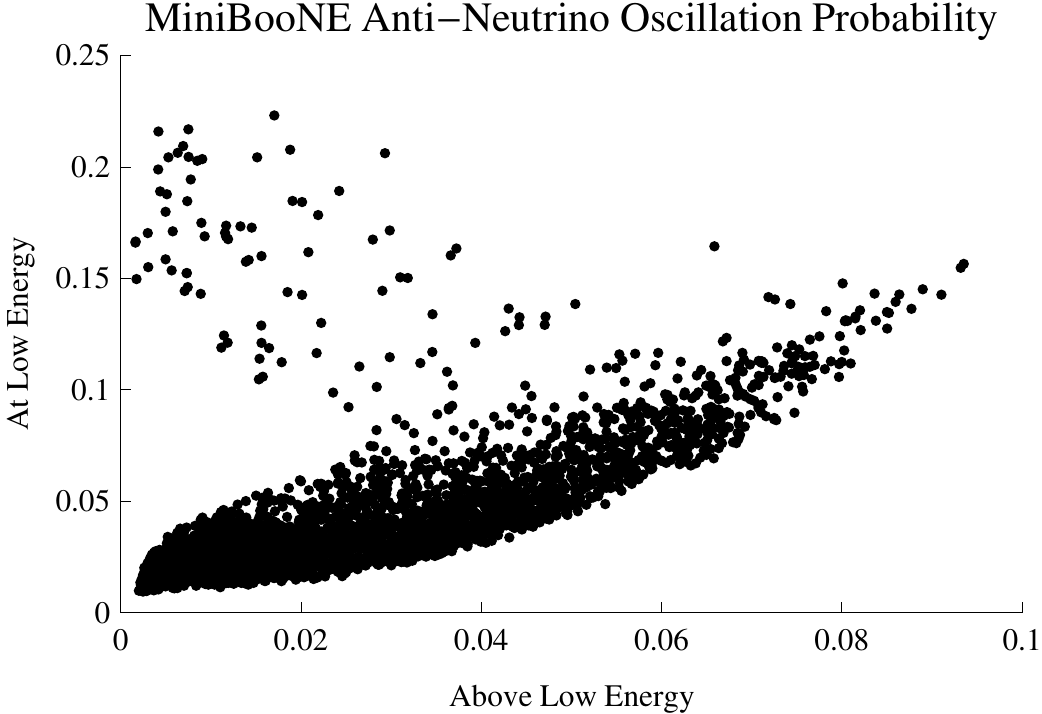}
\caption{The MiniBooNE anti-neutrino oscillation probability in the   region from 200 to 475 MeV vs. the region above 475 MeV}
\label{MBatoMBaloE}
\end{center}
\end{figure}

Due to the resonance in anti-neutrinos in  or just below the  MiniBooNE energy range, the anti-neutrino oscillation probability is consistently much larger than the neutrino oscillation probability over the energy range of the experiment.  We can express this quantitatively over the fit points by the ratio between the anti-neutrino and neutrino oscillation probabilities.  In each energy region, the minimum ratio is above 3; in Figure 4, we give the cumulative distribution of the fraction of fit points with an anti-neutrino to neutrino oscillation probability ratio above a fixed value.

$\\$
\begin{figure}[h!]
\begin{center}
\includegraphics[width=8.0cm]{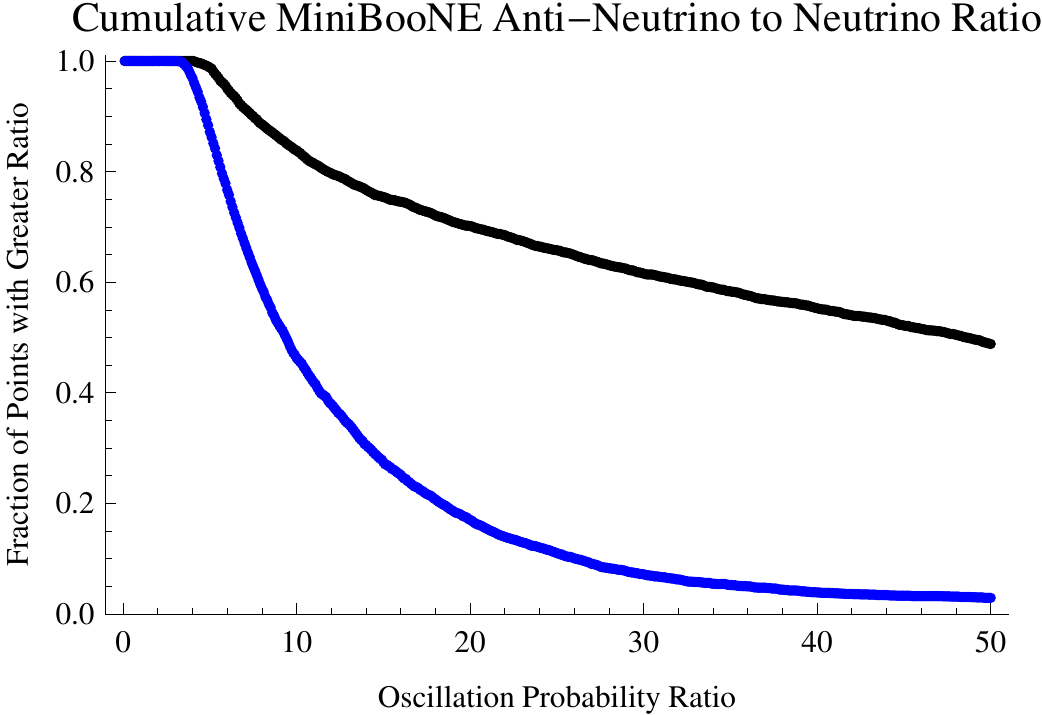}
\caption{The cumulative distribution for ratios of anti-neutrino to neutrino oscillation probability, below (black, upper curve) and above (blue or   gray, lower curve) 475 MeV energy.  For a fixed oscillation probability ratio R, the vertical axis is the fraction of fit points with ratio greater than R}
\label{MBa2MBcumulative}
\end{center}
\end{figure}

Our analysis leads us to make a quantitative prediction for MiniBooNE: anti-neutrino oscillations are likely to be  observable at the experiment.  The experimental systematic uncertainty may become relevant for   for fit points that predict particularly low values of the anti-neutrino oscillation probability in both energy regions, but the majority of fit points correspond to anti-neutrino oscillation probabilities that are large enough to be free of this concern, as shown in Figure 3.

\section{Summary}
We have shown at a model  with three sterile neutrinos and a light $B-L$ gauge boson can accommodate the results of both LSND and MiniBooNE, which are in direct conflict under a standard oscillation model.  The effects of the light vector boson  with weak gauge coupling can be made to fit experimental limits, while having a significant effect on short baseline neutrino oscillations.  We have shown agreement with  other neutrino oscillation experiments besides LSND and MiniBooNE, and made a quantitative prediction that anti-neutrino oscillations  at MiniBooNE are much larger than  neutrino oscillations, and, with sufficient anti-neutrino data, are likely to be visible.

 \section*{Acknowledgments} We would like to acknowledge useful conversations with Neal Weiner, Bill Louis, Eduardo Mass\'o, Michele Maltoni, Maxim Pospelov and Andrea Donini. This work  was partially
supported by the DOE under contract DE-FGO3-96-ER4095.  The work of Jonathan Walsh was supported by a LHC Theory Initiative Graduate Fellowship.  Ann Nelson would like to acknowledge partial sabbatical support from the Ministerio de Educaci\'on y Ciencia of Spain, partial sabbatical support from the University of Washington,  and the hospitality of the Insituto de F\'isica Te\'orica UAM/CSIC at the Universidad Aut\'onoma de Madrid.
 
 \bibliographystyle{apsrev}

\end{document}